\documentclass[journal]{IEEEtran}
\usepackage{cite}
\ifCLASSINFOpdf 
\usepackage[pdftex]{graphicx}
\else
\usepackage[dvips]{graphicx}
\fi

\usepackage[cmex10]{amsmath}
\usepackage{algorithmic}
\usepackage[ruled,norelsize]{algorithm2e}

\usepackage{array}
\usepackage{mdwmath}
\usepackage{mdwtab}
\usepackage{todonotes}
\usepackage{adjustbox}

\usepackage{eqparbox}
\ifCLASSOPTIONcompsoc
\usepackage[caption=false,font=normalsize,labelfont=sf,textfont=sf]{subfig}
\else
\usepackage[caption=false,font=footnotesize]{subfig}
\fi
\usepackage{fixltx2e}
\usepackage{url}
\usepackage{amsfonts}
\usepackage{amssymb}
\usepackage{amsthm}
\usepackage{bm}
\usepackage[ruled]{algorithm2e}
\usepackage[utf8]{inputenc}
\usepackage{booktabs}
\usepackage{url}
\usepackage{cite}
\usepackage{flushend}
\usepackage{subfloat}
\usepackage{subfig}
\usepackage{caption}
\usepackage{epstopdf}

\newcounter{tempEquationCounter} 
\newcounter{thisEquationNumber}

\renewenvironment{IEEEbiography}[1]
  {\IEEEbiographynophoto{#1}}
  {\endIEEEbiographynophoto}

\begin{document}
%\title{Toward Multi-Service Edge-Intelligence Paradigm: Prediction-Communication Co-design for Time-Critical Control over Wireless}
\title{Toward Multi-Service Edge-Intelligence Paradigm: Temporal-Adaptive Prediction for Time-Critical Control over Wireless}
%\title{Multi-Service Edge-Intelligence: A New Paradigm for Time-Critical Control over Wireless}


\author{Adnan~Aijaz,~\IEEEmembership{Senior~Member,~IEEE},
		Nan~Jiang,~\IEEEmembership{Member,~IEEE}, and
		Aftab~Khan
		%Nan~Jiang
\vspace{-0.4cm}
        %Mahesh~Sooriyabandara,
        %and~Usman~Raza
        %and~A.~Hamid~Aghvami,~\IEEEmembership{Fellow,~IEEE}% <-this % stops a space
\thanks{The authors are with the Bristol Research and Innovation Laboratory, Toshiba Europe Ltd., Bristol, BS1 4ND, United Kingdom. Contact e-mail: adnan.aijaz@toshiba-bril.com}}
\markboth{IEEE Internet of Things Magazine -- Accepted for Publication}%
{Shell \MakeLowercase{\textit{et al.}}: Bare Demo of IEEEtran.cls for Journals}


\maketitle
\begin{abstract}
\boldmath
Time-critical control applications typically pose stringent connectivity requirements for communication networks. The imperfections associated with the wireless medium such as packet losses, synchronization errors, and varying delays have a detrimental effect on performance of real-time control, often with safety implications. \textcolor{black}{This paper introduces \emph{multi-service edge-intelligence} as a new paradigm for realizing time-critical control over wireless. It presents the concept of multi-service edge-intelligence which revolves around tight integration of wireless access, edge-computing and machine learning techniques, in order to provide stability guarantees under wireless imperfections. The paper articulates some of the key system design aspects of multi-service edge-intelligence. It also presents a \emph{temporal-adaptive prediction} technique to cope with dynamically changing wireless environments.} It provides performance results in a robotic teleoperation scenario. Finally, it discusses some open research and design challenges for multi-service edge-intelligence. 
%\vspace{9pt}
%\emph{Index Terms}---Cognitive Radio Ad-Hoc Networks, Cooperative Routing, Spectrum %Aggregation
\vspace{-1.5em}
\end{abstract}

% \fi
\IEEEpeerreviewmaketitle

\section{Introduction}
\IEEEPARstart{R}{eal}-time control systems (RTCSs) underpin critical applications across a range of industrial domains including manufacturing, oil and gas, energy distribution, nuclear decommissioning, and space exploration. RTCSs are time-critical in nature and typically involve feedback (closed-loop) control wherein spatially-distributed controllers, sensors, and actuators exchange command and feedback messages over a communication medium. With recent trends toward industrial Internet-of-Things (IoT) and Industry 5.0, new applications for RTCSs are emerging including multi-robot formation control, human-robot collaboration, and multi-modal teleoperation \cite{TI_PIEEE}.

Performance requirements of RTCSs can be quite stringent, especially in terms of timeliness\footnote{Timeliness refers to deterministic latency guarantees which implies that latency arising from communication must have very low variance, i.e., minimal jitter, between consecutive cycles. } and ultra-grade reliability and responsiveness \cite{3gpp_22_104}. Hence, wired solutions based on Fieldbus and Ethernet technologies are dominant in industrial environments. Wireless technologies provide a low-cost alternative with additional benefits of flexibility and mobility support. \textcolor{black}{However, the inherent uncertainty associated with the wireless medium manifests in the form of latency variations, packet losses, and time synchronization errors. Such imperfections are detrimental to the stability of RTCSs and lead of system outage or loss of transparency, often with safety implications.}

\begin{figure*}
\begin{center}
\includegraphics[scale=0.5]{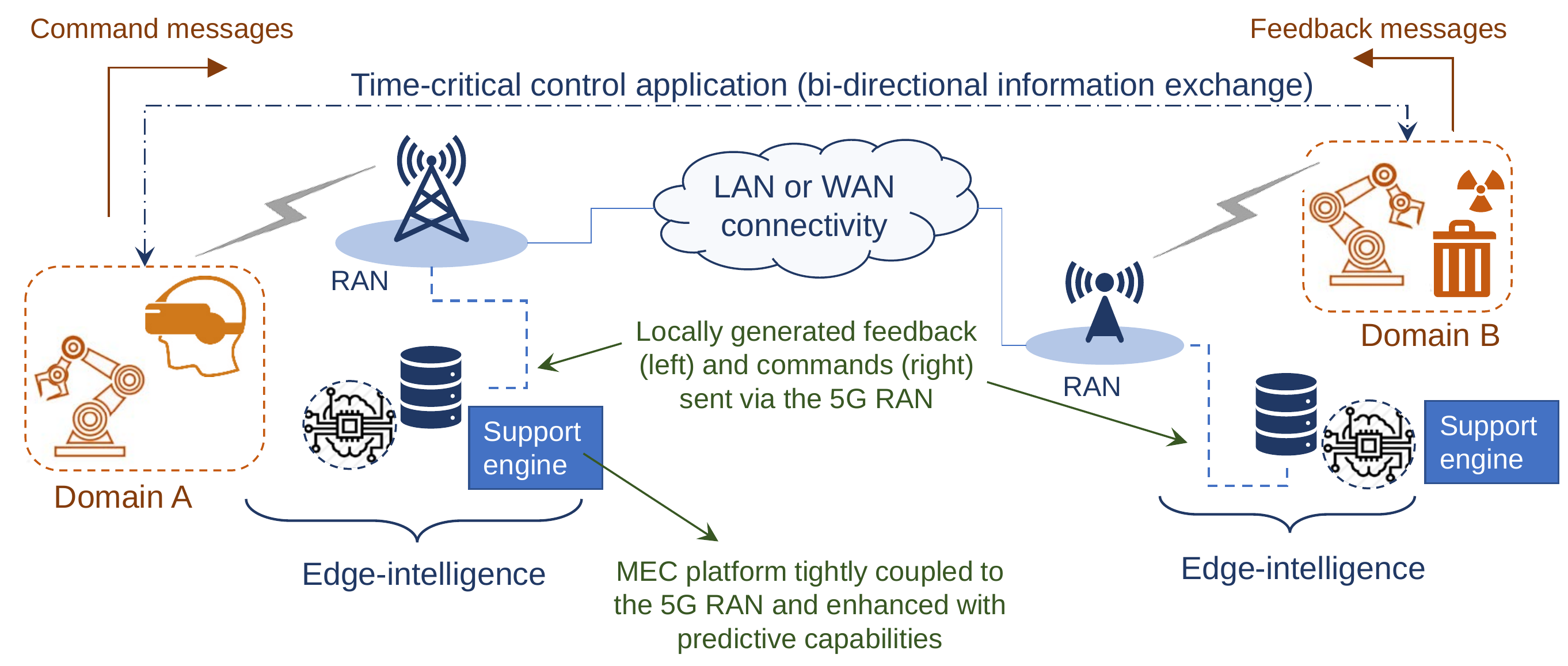}
\caption{\textcolor{black}{Illustration of the multi-service edge-intelligence concept.} }
\label{concept}
\end{center}
\vspace{-2em}
\end{figure*}

Untethered RTCSs based on wireless technologies are still in infancy. Recently, the viability of closed-loop control over wireless has been demonstrated \cite{GALLOP_journal}; however, such solutions are limited to local environments. \textcolor{black}{The fifth-generation (5G) mobile/wireless technology fulfils the latency requirements of most real-time applications; however, it does not provide the much-needed determinism required by RTCS. 
%performance requirements of RTCSs becoming challenging to meet due to various reasons. 
First, the performance experienced by a user varies as a function of distance from the base station; hence ``anytime and everywhere" guarantees are hard to provide. Second, even though 5G natively supports ultra-reliable low-latency communication (uRLLC), realizing feedback control requires enhancements at different layers of the air-interface to guarantee minimal jitter for bi-directional exchange. Third, over-the-air time synchronization techniques for external grandmasters necessitate frequent signaling \cite{perf_OTA_time_synch} which cannot always be guaranteed as a single control-plane is often shared across different user-planes. Last but not least, increased separation between controllers and sensors/actuators necessitates information exchange over interconnected systems (e.g., via the Internet) where latency cannot be easily guaranteed to additional communication and computation factors.}

\textcolor{black}{Conventional paradigms for realizing control over wireless can be classified into (a) \emph{control-aware wireless design}  and (b) \emph{wireless-aware control design}. The former aims to design high-performance wireless protocols \cite{TI_PIEEE} for meeting real-time requirements, e.g., the IO-Link Wireless protocol \cite{IOLW}. The latter focuses on designing control algorithms and architectures to cope with communication uncertainties, e.g., passivity-based controllers for robotic applications \cite{cont_to1,cont_to2}.}

\textcolor{black}{This paper introduces a new multi-service edge-intelligence framework for guaranteed stability of RTCSs in the presence of imperfections associated with the wireless medium such as packet losses, time synchronization errors, and latency variations. Multi-service edge-intelligence is different from conventional paradigms for realizing control over wireless. It revolves around tight coupling of wireless access, edge-computing, and predictive techniques, \emph{without specially designed wireless protocols or application-specific control algorithms.} It empowers any kind of wireless network to handle real-time control. It unlocks the potential of real-time control at scale for industry and society without requiring specialized robotics hardware and devices with proprietary interfaces. } 
To this end, the key contributions of this work are highlighted as follows.
\begin{itemize}
\item We provide a holistic perspective on multi-service edge-intelligence framework with some of the key system design aspects (Section II). 
\item As part of the multi-service edge-intelligence framework, we present a \textcolor{black}{temporal-adaptive prediction} technique (Section III).
\item We conduct performance evaluation through realistic simulations, aided by a practical dataset, in a human-centric robotic manipulation scenario (Section IV).
\item We discuss some of the key open research and design challenges for multi-service edge-intelligence (Section V). 
\end{itemize}

\section{Multi-service Edge Intelligence} \label{fw}
\subsection{The Framework}
\textcolor{black}{The multi-service edge-intelligence framework adopts a co-design approach for (a) multi-access edge-computing (MEC), (b) artificial intelligence (AI) and machine learning (ML) techniques\footnote{\textcolor{black}{We use the term ML throughout the paper as it is a sub-area of AI. }}, and (c) wireless access system (5G RAN). This co-design has two integral components: (i) tight integration from architectural aspects, and (ii) joint optimization from protocol and algorithmic aspects. } The framework is expected to be generic to cater for any kind of control application (running as a service) over a communication system. \textcolor{black}{The main motivation for the framework is to guarantee stability of time-critical control applications in the presence of wireless imperfections.} However, it also enables the perception of real-time connectivity in human-centric control applications and overcomes the physical limitations arising due to bottlenecks in integrated systems and the finite speed of light. 

The concept of multi-service edge-intelligence is illustrated in Fig. \ref{concept} which depicts an immersive teleoperation scenario in an industrial environment (e.g., for nuclear waste decommissioning). A time-critical control application is running over a communication network. The first domain (Domain A or the master domain) generates control commands, e.g., a human operator interacting with a master robot. The second domain (Domain B or the slave domain) receives control commands, performs actuation, e.g., a slave robot handling a task, and sends feedback to the first domain. The two domains are wirelessly-connected, e.g., via 5G RANs which are connected via either local area or wide area networking technologies.

Multi-service edge-intelligence is realized via support engines\footnote{High-level concept of support engines has been discussed in the recent IEEE P1918.1 standard as well \cite{GC_TI_arch}; however, it has not been explored from a holistic view covering tight integration of different system elements as described in this work. Moreover, edge-intelligence techniques have received little attention from a protocol design perspective.} running in proximity of the domains exchanging time-critical control information as shown in Fig. \ref{concept}. \textcolor{black}{A support engine is an edge-computing platform, providing computational and storage resources, and tightly coupled to the 5G system. It is equipped with model training and inference capabilities using ML techniques. For multi-service edge-intelligence, support engines provide crucial predictive functionalities for locally generating command/feedback signals and their timely delivery in case the actual signals are lost or delayed due to imperfections of the communication system. Note that the support engines are running in proximity of both domains. For the slave domain, the role of support engine is to predict the command messages, whereas for the master domain, it predicts the feedback messages. The predicted information is delivered to respective domain through the 5G RAN. }

\textcolor{black}{The co-design approach underpinning multi-service edge-intelligence framework requires various design considerations from a system-level perspective. These are discussed below from the perspective of challenges as well as potential solutions.}

\subsection{\textcolor{black}{System-level Design: The Tightly-Coupled MEC Challenge}}
\begin{figure}
\begin{center}
\includegraphics[width=\columnwidth]{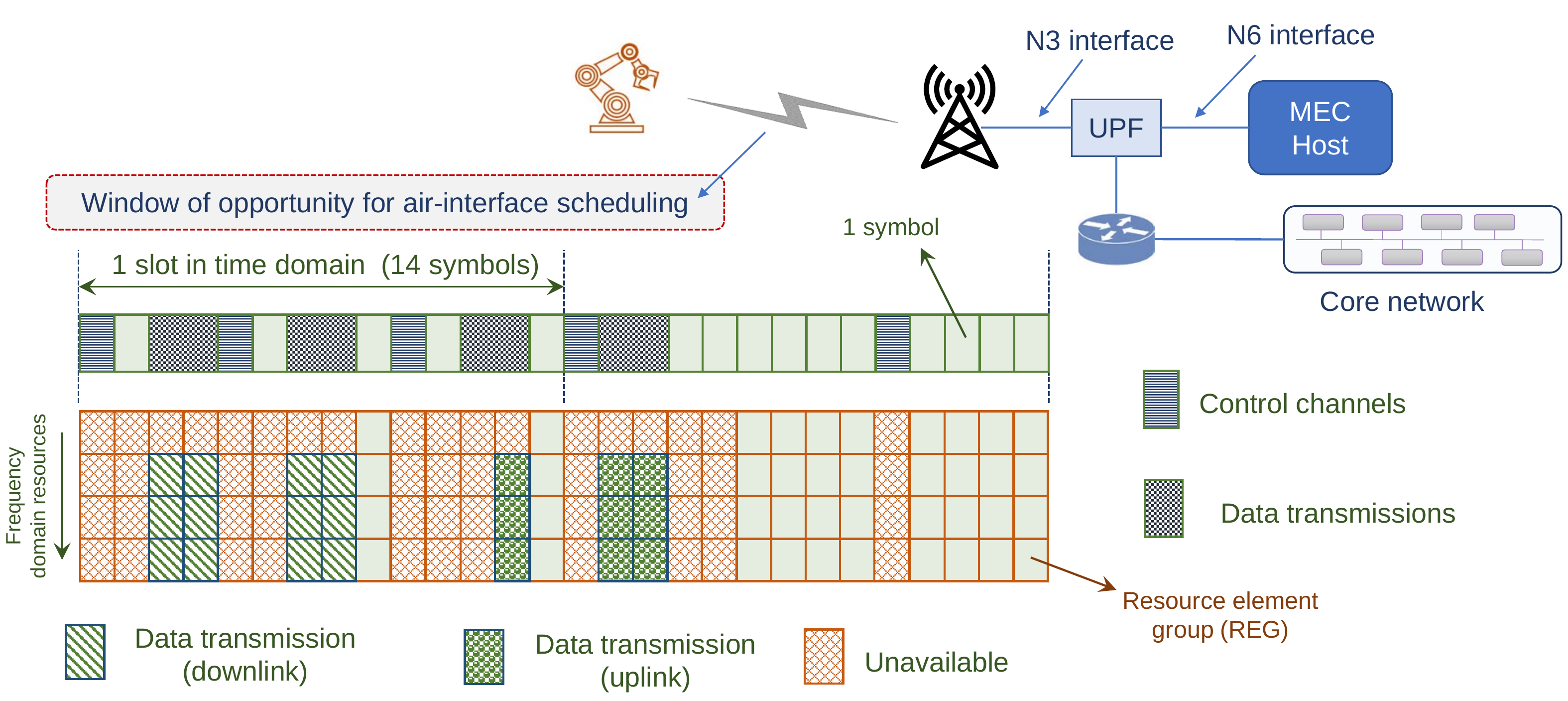}
\caption{\textcolor{black}{MEC deployment in 5G for multi-service edge-intelligence.} }
\label{5G-MEC-coupling}
\end{center}
\vspace{-1em}
\end{figure}

A key system design aspect for multi-service edge-intelligence is tight coupling of 5G and MEC systems. \textcolor{black}{This requires architectural-level as well as protocol-level enhancements. From an architectural perspective, the MEC deployment must be in close proximity of the wirelessly-connected control/actuating edges.} The 3GPP service-based architecture provides flexibility in deploying the user-plane function (UPF). Therefore, the UPF and the MEC system can be co-located with the 5G RAN for realizing the multi-service edge-intelligence framework. In such a scenario, the MEC system will be deployed in a data network external to the 5G system, i.e., on the N6 reference point, as illustrated in Fig. \ref{5G-MEC-coupling}. \textcolor{black}{One example of connected 5G and MEC system is the Aether platform (https://opennetworking.org/aether/). }

\textcolor{black}{The protocol-level challenge is further split into traffic steering, radio resource allocation, and synchronized operation aspects.} The traffic steering capability to/from the MEC becomes particularly important for model training as well as for delivery of the predicted information to the edge. In an integrated 5G-MEC system, the traffic steering capability becomes the responsibility of the UPF; however, MEC system can influence traffic steering through interaction with the 5G network functions (policy control function,  application function, etc.) as various new 5G functionalities support enhanced integration with the MEC system \cite{MEC_5G_white_paper}. Selective traffic steering can be achieved through configuring the UPF with downlink and uplink classifiers. 

\textcolor{black}{The resource allocation technique in the 5G RAN must ensure timely and reliable delivery of the \emph{actual} as well as the \emph{predicted} control/feedback information for actuation as well as necessary information for the predictors.} This implies a window of opportunity for the air-interface during which downlink/uplink resource allocation must take place. This can be achieved through either proactive or reactive resource allocation, potentially with joint allocation for downlink and uplink. 

The radio resource allocation aspects are directly influenced by the level of synchronization between the MEC and the 5G RAN. A loosely-synchronized integrated system will require an event-triggered approach for resource allocation. However, a tightly-synchronized system, underpinned by a time-sensitive networking (TSN) interface connecting MEC and RAN, paves the way for time-triggered resource allocation.

\subsection{\textcolor{black}{System-level Design: The ML at the Edge Challenge}}
%\subsection{System-level Design Aspects: The MEC+AI Challenge}
\textcolor{black}{Another crucial system design aspect is the integration of MEC system with ML techniques. This is important for realizing predictive intelligence as an edge application. }  
The MEC framework and reference architecture, as defined in ETSI \cite{MEC_ETSI}, enables implementation of MEC applications as software-only entities running on top of a virtualization infrastructure. Tight coupling of MEC and ML implies that the MEC system is utilized as a platform for running edge-intelligence as a MEC application. \textcolor{black}{The main functional blocks for edge-intelligence as a MEC application include a predictive engine, a training module, an event handler, and a system controller for traffic routing and allocation of computation resources.} These functional blocks must be mapped onto the MEC reference architecture and reference points, particularly in terms of interaction with (a) the MEC platform (via the Mp reference points), (b) host-level and system-level management (via the Mm reference points), and (c) external 3GPP system (via the Mx reference points).

\textcolor{black}{An alternative solution is provided in the form of the O-RAN architecture (https://www.o-ran.org/), which is built on the principles of openness and intelligence, and shares some conceptual similarities with the MEC architecture.} Integration of MEC and O-RAN architectures is beneficial from various perspectives \cite{integ_ORAN_MEC}. O-RAN brings intelligence at the edge of the RAN in the form of RAN intelligent controllers (RICs) \cite{O-RAN_arch}, thereby paving the way for native integration of ML capabilities in 5G networks. In particular, the non-real-time RIC (non-RT RIC) supports ML workflow including model training and policy-based guidance of applications/services.

\subsection{\textcolor{black}{System-level Design: The ML for Wireless Challenge}}
%\subsection{System-level Design Aspects: The AI+Wireless Challenge}
\textcolor{black}{Optimizing ML techniques for the peculiarities of wireless environments plays an important role in multi-service edge-intelligence.}  Conventional ML prediction functions (e.g., exponential smoothing) provide fast convergence and consume less computational resources. However, such predictors suffer from significant reduction in accuracy when the observations are decreasing (e.g., due to wireless link outage), and therefore only suitable for predicting over a single slot. On the other hand, deep learning predictors offer higher accuracy and can be used for multi-slot prediction; however, these are intensive in terms of computational resource requirements and their execution time can be high. \textcolor{black}{Multi-slot prediction becomes particularly important considering exchange of command/feedback messages with multiple degrees-of-freedom (DoF) in teleoperation applications and also to mitigate the impact of burst errors in wireless environments. It is important to dynamically adapt the prediction horizon, by utilizing different types of predictors, as per the wireless channel; however, it entails changing the underlying model and the training requirements. In the next section, we describe a prediction technique for addressing this challenge. }
%On the other hand, control applications like teleoperation often involve exchange of command/feedback signals with multiple degrees-of-freedom (DoF) having varying levels of tolerance to wireless imperfections. Hence, it is desirable to adapt different types of predictors as per the wireless environment. 

\section{\textcolor{black}{Temporal-Adaptive Prediction Technique}}
%\section{Prediction-Communication Co-design}
\textcolor{black}{To provide a generic service-agnostic framework, and to cope with dynamically changing wireless environments, we design a \emph{temporal-adaptive prediction} (TAP) technique where different prediction models, with different capabilities and complexities, are deployed to run in parallel.} Due to its lower complexity, the prediction horizon of the short-term predictor is limited, i.e., it only predicts command/feedback signals in the near future. The long-term predictor, with higher prediction capabilities, has a broader prediction horizon. It can support operation at the edges until new command/feedback signals are successfully received. Note that previous prediction results are discarded once new command/feedback signals are received.

\begin{figure}
\begin{center}
\includegraphics[width=\columnwidth]{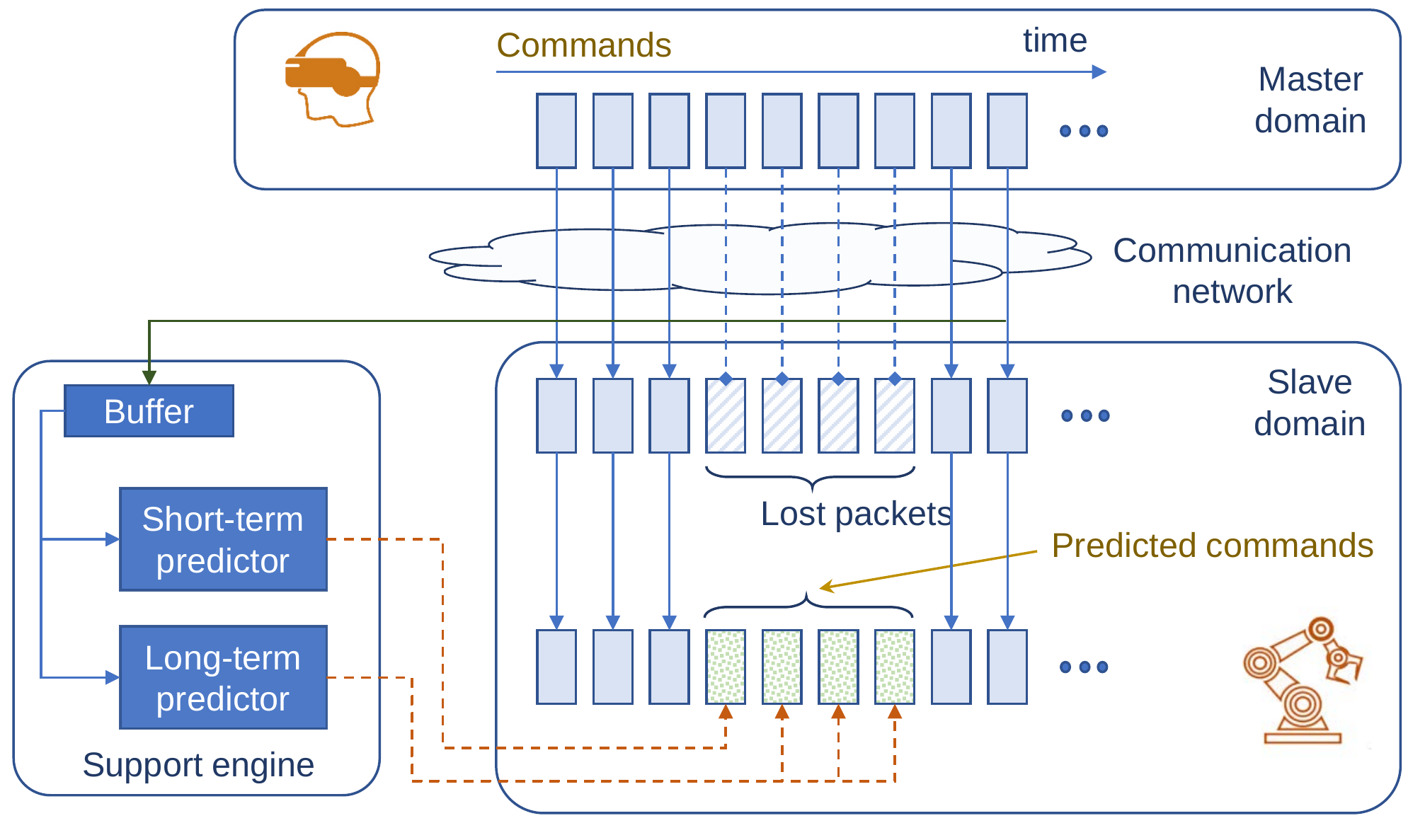}
\caption{The TAP technique for multi-service edge-intelligence. }
\label{tap_concept}
\end{center}
\vspace{-1em}
\end{figure}

\textcolor{black}{The TAP technique is illustrated in Fig. \ref{tap_concept} which depicts a time-slotted model for communication between the master and slave domains in a teleoperation scenario. In each timeslot, the master domain generates a new command message which is transmitted to the slave domain via the communication network. The figure also shows a support engine, containing a packet buffer and the two predictors, for providing predictive capabilities in proximity of the slave domain.  As shown, the short-term predictor is solely responsible for predicting the lost commands in the fourth timeslot. The long-term predictor supports the slave domain until the new commands are successfully received, i.e., it provides multi-slot prediction and predicts commands in fifth, sixth, and seventh timeslots.  }

\textcolor{black}{The TAP technique operates as follows. At any given time \(t\), once a new command (or feedback) signal/message \(O^t\) has been received, the short-term as well as the long-term predictors are triggered to predict the forthcoming signals.} This is achieved by gathering a set of signals \(W^t = [O^{t-\phi}, O^{t-\phi+1}, \hdots, O^{t-1}, O^t] \) from the buffer and feeding into both predictors. At the same time, the slave robot executes actions according to the newly received command \(O^t\). In the next slot, the predictors are re-initialized if a new command has been successfully received.  Otherwise, it executes actions according to the commands generated by predictors.

The short-term predictor utilizes time series analysis tools like vector autoregressive (VAR) and auto regressive integrated
moving average (ARIMA). These predictors exploit the inherent natural structure in the time series data by learning through partial historical data via unsupervised clustering mechanisms.

\begin{figure}
\begin{center}
\includegraphics[scale=0.3]{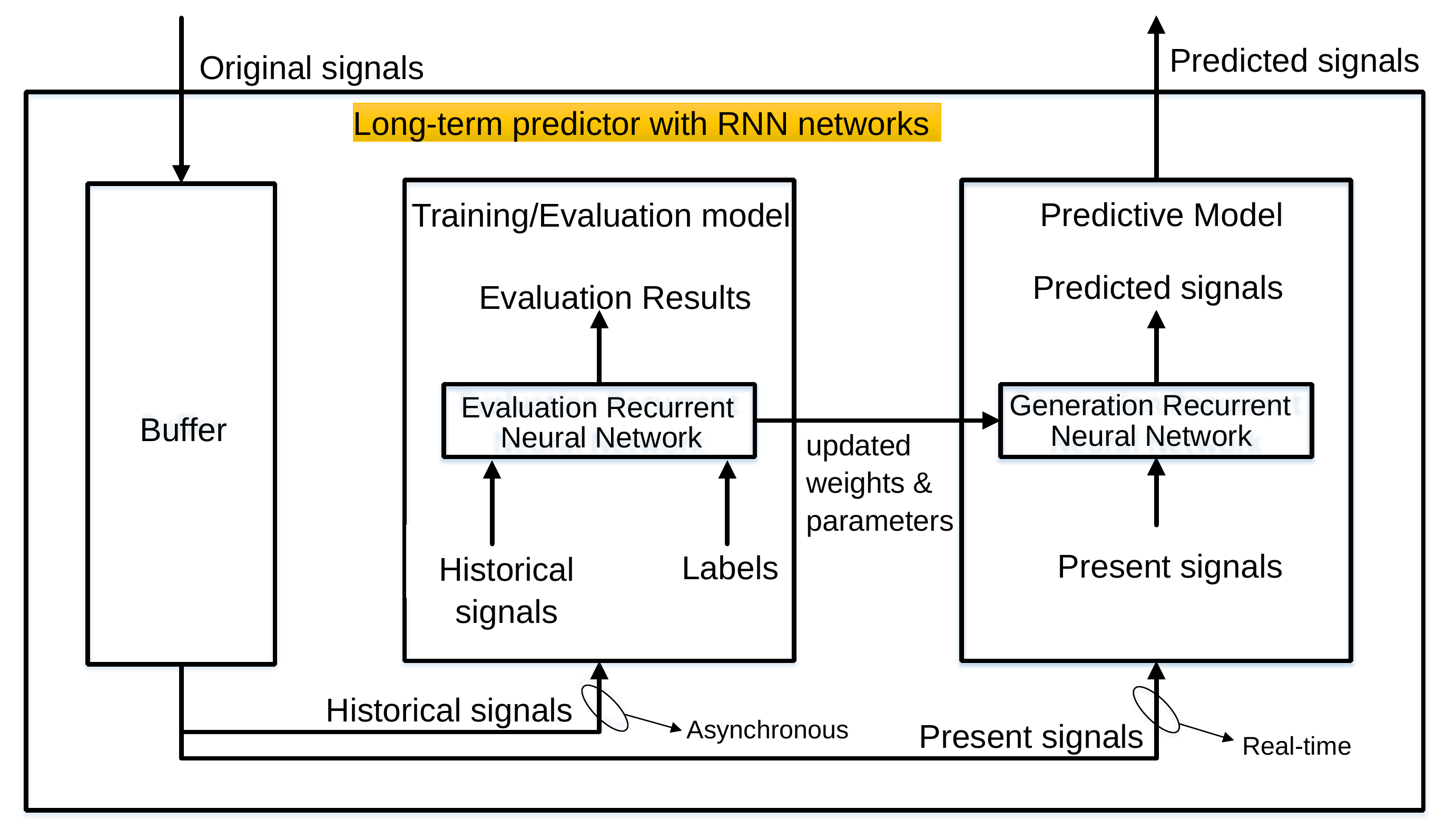}\quad
\includegraphics[scale=0.38]{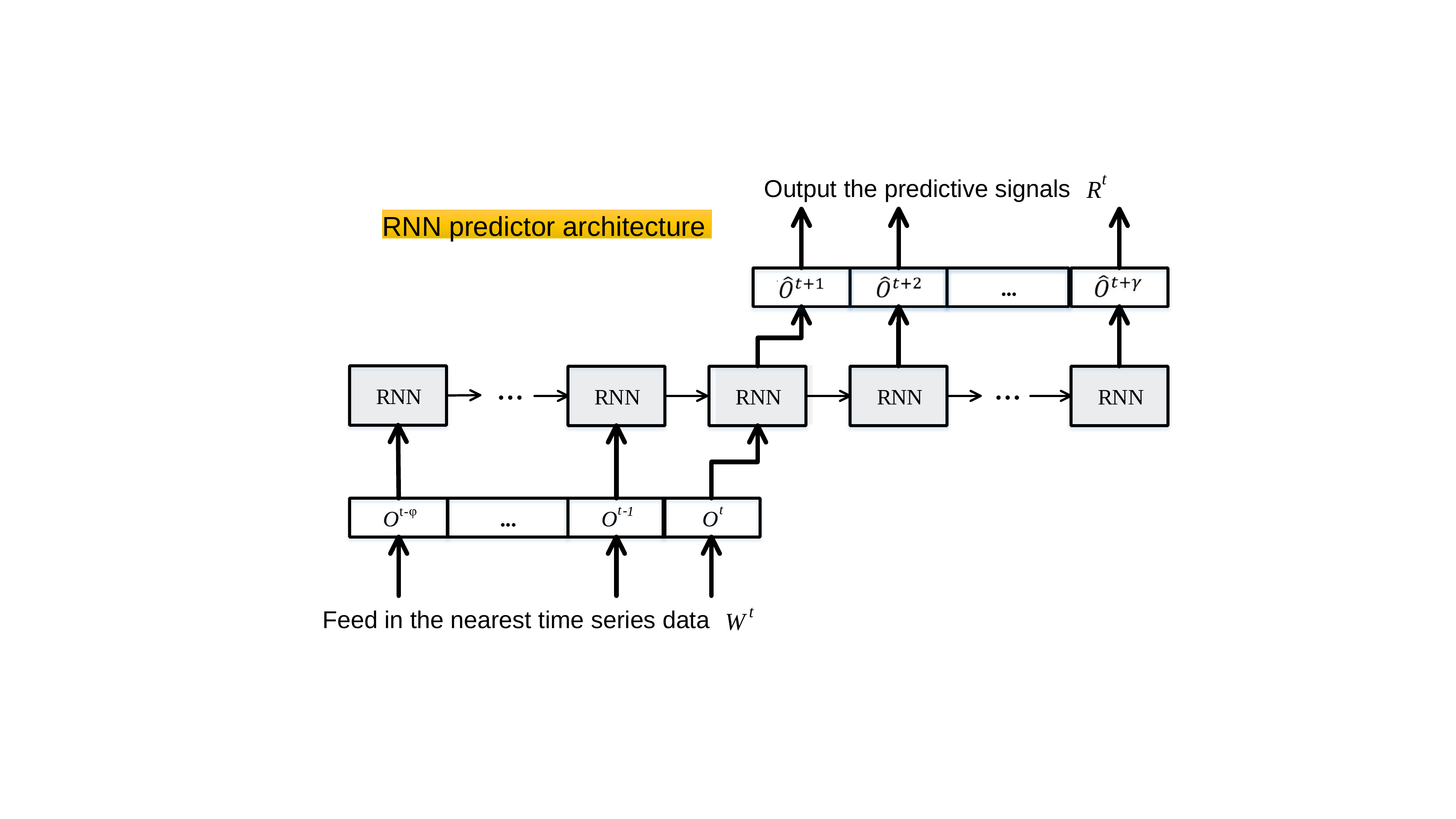}
\caption{Illustration of the long-term predictor (top) and the RNN predictor architecture (bottom).  }
\label{long_term}
\end{center}
\vspace{-1em}
\end{figure}

The long-term predictor, which is illustrated in Fig. \ref{long_term}, utilizes a supervised learning model, which aims at accurately capturing the temporal dynamics of the long-memory time series data and the complex correlation among multiple DoF. We adopt a recurrent neural network (RNN) based on the gated recurrent units (GRU) architecture.  \textcolor{black}{At any given slot \(t\), the original command (or feedback) signal $O^t$ is received by the predictive learning model stored in the buffer. The generation system is a well-trained RNN located in the support engine. The generation RNN consist of several layers of a RNN cell where the last RNN layer is connected to an output layer consisting of several activation functions. Once the buffer receives a new original signal, it immediately inputs the nearest time series data \(W^t\) into the generation system such that
\(W^t = [O^{t-\phi}, O^{t-\phi+1},\hdots, O^{t-1}, O^t]\). During the  feedforwarding progress, the RNN is progressively fed the previous signals $O^{t-\phi}, O^{t-\phi+1},\hdots, O^{t-1},O^t$, and it then progressively produces the vector of predicted signals
\(R^t = [\hat{O}^{t+1},\hat{O}^{t+2},\hdots,\hat{O}^{t+\gamma}] \). }

The predictive horizon is limited by the factor $\gamma$. As execution delay of the RNN predictor can be longer than one slot, the actuator in physical environment always relies on the short-term predictor at the beginning of a prediction cycle, and then it  switches to the long-term predictor once the initial prediction procedure is complete. \textcolor{black}{Finding the optimal number of timeslots for the operation of short-term and long-term predictors is left as part of any future work. }

In some scenarios, the training procedure mainly occurs offline through prior data from simulation or training. In these cases, a training/evaluation system is still employed in the support engine for on-the-fly fine-tuning.  The evaluation RNN has the same architecture as the generation RNN. The training frequency depends on the requirement, which is not synchronized with generation cycle. Once required, the buffer randomly picks a batch of \textcolor{black}{normalized training samples}, each sample at slot $\tau$ contains an input matrix (historical signals)
\( W^\tau = [O^{\tau-\phi}, O^{\tau-\phi+1},\hdots, O^{\tau-1}, O^\tau]\), and and a label matrix
\(  L^\tau = [O^{\tau+1},O^{\tau+2},\hdots,O^{\tau+\gamma}] \).

These training samples are fed into the evaluation RNN to calculate an average loss for the gradient descent. For instance, in one training method, the loss can be obtained by calculating the mean squared error between output results and labels. In both generation and training/evaluation systems, the prediction accuracy is measured by a method, e.g., calculating average absolute errors between predictive results and labels. Once the performance of the training/evaluation systems outperforms that of the generation one, the obtained gradient during training will be shared between them for updating the weights.

Note that the training procedure can also be conducted in an online manner without any pre-training.  At the beginning, the system executes the single prediction mode, where only the short-term predictor is triggered. The training/evaluation system in the long-term predictor is initialized by randomly generating weights. The training/evaluation system is trained using the samples in the buffer. Note that the training is not synchronized with the generation cycle, which will not occupy computational resource during generation procedure. In this scenario, the prediction accuracy of the short-term predictor is also measured. After every training epoch, the long-term predictor will be evaluated by comparing its prediction accuracy with the short-term predictor. Once the prediction accuracy of the long-term predictor outperforms the short-term predictor, the weights of the training/evaluation system are shared with the generation system, and the TAP mode will be triggered.

The size of control commands can be smaller than the maximum payload carrying capacity of a packet on the air-interface. In such scenarios, a command-bundling transmission (CBT) technique can be adopted which improves resource utilization while providing additional features for prediction. We assume that multiple successive commands can be included in the same packet. The received packet may include out-of-date control signals, which will be used as the extra input for the predictors. \textcolor{black}{Let, \(f_s\) denote the sampling rate of the controller and \(f_t\) denote the maximum transmission rate of the network. Then, at each transmission interval, \(\mu=\lceil f_s/f_t \rceil \) can be included into one packet and transmitted via the network.} Considering a single packet is
received by the predictor under the timing constraint, only the
\(\mu\)th command will be directly applied to the actuator, while \(\mu-1\) commands are outdated and these will only be
utilized for future prediction.

\section{Performance Evaluation}
\begin{figure}
    \centering
    \includegraphics[width=\columnwidth]{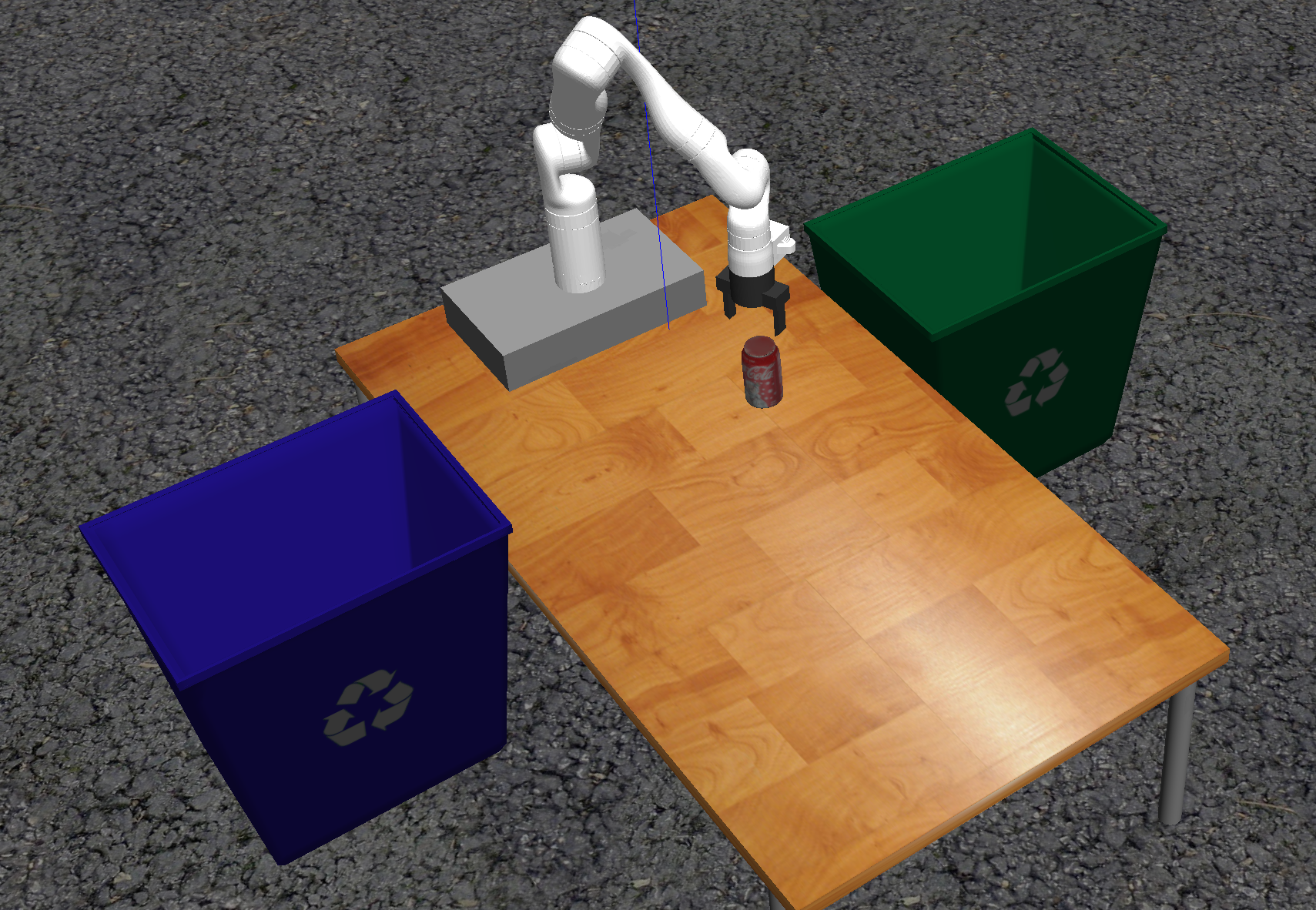}
    \caption{Simulation environment for the robotic manipulation experiment.}
    \label{fig:sim_env}
    \vspace{-1em}
\end{figure}

We simulate a robotic manipulation scenario, using a customized simulator written in MATLAB, Gazebo, and Python, where a remotely-controlled (by a controller) robotic arm with 7 DoF aims at picking an object (a coke can in this case) and putting it into a container as shown in Fig.  \ref{fig:sim_env}. 
The controller has prior knowledge of the location of the target object and it is responsible for calculating all required control signals of the manipulator. 
The operation is time-slotted and the control signals are time-varied sequences, where, at a single slot, a command matrix includes the information of planned positions, speeds, and accelerations of each DoF. 
At each slot, the controller transmits the current command matrix to the manipulator and waits for the feedback matrix. 
The controller transmits the next command matrix only when it confirms the manipulator arriving at the expected posture according to the received feedback matrix. 
We assume that the network link for transmitting commands is not perfect and it introduces a random latency in communication. 
We model the latency by a Normal distribution with a mean of 10 millisecond (msec) and a variance of  20 msec. 
At each slot, if the packet with the current command matrix was successfully received within the delay constraint, the manipulator executes actions according to the received commands. 
If this packet was lost or received with a delay, we consider two different control strategies: 1) benchmark strategy (non-predictive model) where the control happens  according to the last available successful command, and \textcolor{black}{2) predictive method where the control takes place according to the TAP-based predictive model, where the VAR method is used as the short-term predictor.}

%\begin{figure*}
%    \centering
%    \begin{tabular}{cc}
%    \adjustbox{valign=t}{\subfloat[\label{fig:ae_sig}]{%
%          \includegraphics[width=.49\linewidth,trim={0 9em 0 0}]{r2_updated}}}
%    &      
%    \adjustbox{valign=t}{\begin{tabular}{@{}c@{}}
%    %\subfloat[\label{fig:hand_kin}]{%
%          %\includegraphics[width=.30\linewidth]{hand_kin2.png}} \\
%    \subfloat[\label{fig:tap_perf}]{%
%          \includegraphics[width=.49\linewidth]{tap_perf2.pdf}}
%    \end{tabular}}
%    \end{tabular}
%    \caption{(a) Performance comparison of the absolute error indicator from the non-predictive and the predictive signals, (b) Illustration of kinematics data of human hands signals, and (c) Performance comparison of the average absolute error among TAP, non-predictive method, and classical single-predictive method. }\label{fig:perf_hand_tap}
%  \end{figure*}

The position signals from the source, the non-predictive model, and the predictive model, at each slot \(t\), are denoted by \(P^t_s\), \(P^t_{np}\), and \(P^t_p\), respectively.  \textcolor{black}{Input signals are re-scaled between 0 and 1 as a pre-processing step prior to model training. Such data scaling ensures stabilized training with lower gradient errors as well as reduced convergence time \cite{LeCuBottOrrMull9812}.}
%Figure \ref{fig:pos_sig} plots trajectories of those position signals. 
%We can clearly observe that, for each DoF, the signals from the predictive model is fairly close to the source, while those from the non-predictive model suffer from large inaccuracy. 
The  absolute error (AE) between the signals from the source and the non-predictive model is given by \(E_{np}^t=|P_{np}^t-P_s^t|\), whereas 
the AE between signals from the source and the predictive model is given by \(E_p^t=|P_p^t-P_s^t | \).

%\begin{equation}
%    E_{np}^t=|P_{np}^t-P_s^t|    
%\end{equation}
%%
%and the AE between signals from the source and the predictive model is: 
%
%\begin{equation}
%    E_p^t=|P_p^t-P_s^t |    
%\end{equation}

\begin{figure}
    \centering
      \subfloat[\label{fig:ae_sig}]{%
       \centering
    \includegraphics[width=\columnwidth]{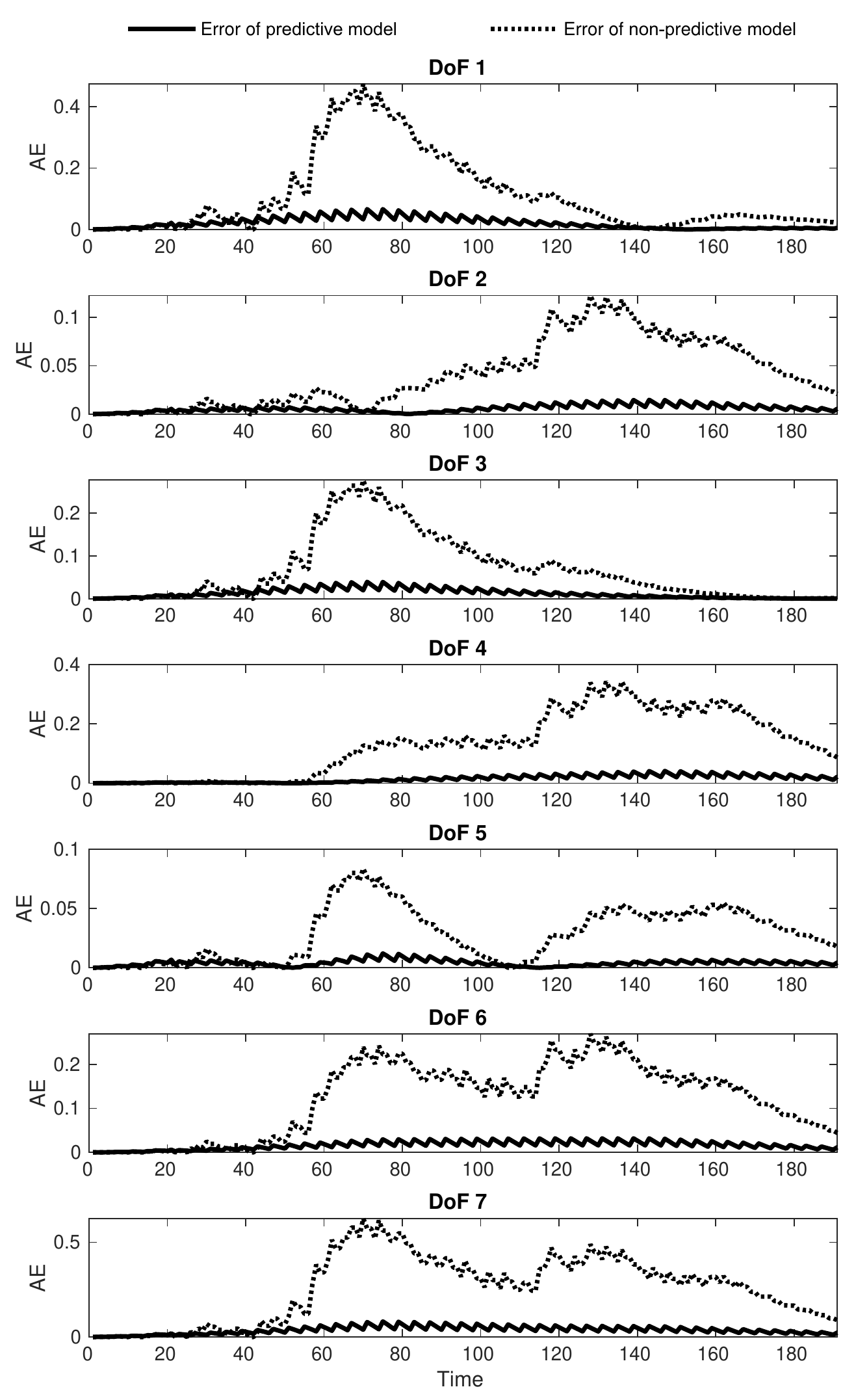}}
    \hfill
      \subfloat[\label{fig:tap_perf}]{%
       \centering
    \includegraphics[width=\columnwidth]{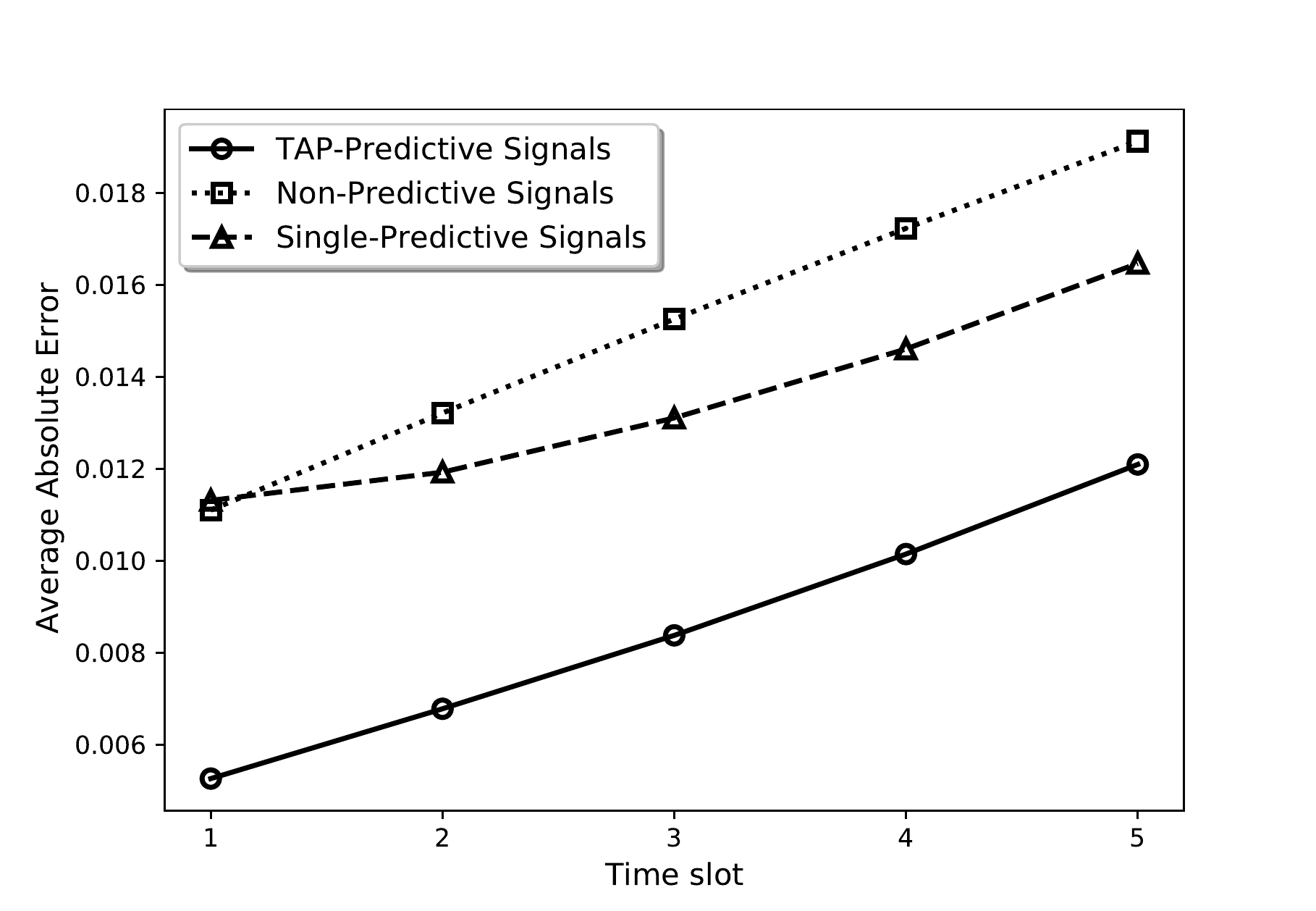}}
    \caption{Performance comparison of \textcolor{black}{(a) the AE from predictive and non-predictive models where the former is based on the TAP technique}, and (b) the TAP technique and its benchmarking. }
    \label{fig:perf_hand_tap}
    \vspace{-1.5em}
\end{figure}

Fig. \ref{fig:ae_sig} shows the AE \textcolor{black}{(in normalized scale)} comparison between the predictive and the non-predictive models. The results reveal that the AE of 
 the predictive model is significantly smaller and it exhibits much lower fluctuations as compared to the non-predictive model. Next, we capture the success probabilities of achieving the pick-and-plance task for different models. We consider two scenarios: a high latency scenario where the mean and variance of latency, as per the Normal distribution, are 10 msec and 20 msec, respectively, and a low latency scenario where the mean and variance are 5 msec and 10 msec, respectively. The results, averaged over 100 iterations, are as follows.
 
\begin{itemize}
\item \textcolor{black}{Predictive model: 0.83 (high latency) \& 0.87 (low latency)} 
\item \textcolor{black}{Non-predictive model: 0.36 (high latency) \& 0.49 (low latency)}
\end{itemize} 

The results show that the predictive model considerably outperforms the non-predictive model in terms of task performance. 

%The success probabilities of achieving picking-and-putting task for the predictive model and the non-predictive model within the high latency (i.e., Normal distributed latency with a mean $0.01 s$ and a variance $0.02$) and the low latency (i.e., Normal distributed latency with a mean $0.005s$ and a variance $0.01$) environments are given in Table \ref{tab:my_label}. 
%Each result is obtained after $100$ iterations. 
%We clearly observe that the predictive model considerably outperforms the non-predicative model.
%
%\begin{table}[t]
%    \centering
%    \caption{Task success probabilities}
%    \begin{tabular}{lcc}
%    \toprule
%    & Predictive mode & Non-Predictive mode \\
%    \midrule
%    High Latency & 83/100 & 36/100  \\
%    Low Latency & 87/100 & 49/100 \\
%    \bottomrule
%    \end{tabular}
%    \label{tab:my_label}
%\end{table}

Next, we conduct performance evaluation of the TAP technique using a practical dataset \cite{dataset}, which collects kinematics data of human hands during the performance of a wide variety of activities of daily living involving feeding and cooking. 
%As shown in Figure \ref{fig:hand_kin}, 
This dataset was recorded with measurements in 18 DoF on each hand (see Fig. 3 in \cite{dataset} for more details). 
For simplicity, we only consider the data from one hand here.

We consider the scenario of remotely controlling a slave robot via transmitting the kinematics data through a wireless network. We assume that  
the transmission rate of the network is limited such that only one command can be transmitted every six slots. 
Therefore, after receiving a new command, the predictor is responsible to predict the commands in the next five slots. \textcolor{black}{Our aim is to create a challenging scenario for the predictor. Besides, transmitting one command every six slots represents the scenario where the frequency of incoming signals is higher than maximum possible frequency of delivering, e.g., in the case of haptic streams which are typically sampled at a very high rate of 1 kHz \cite{TI_PIEEE}.} 

Fig. \ref{fig:tap_perf} plots the average AE between the source and the predictive models, including the proposed TAP technique, the non-predictive method, and \textcolor{black}{classical single-predictive method which only utilizes a single predictor, i.e., the VAR technique}. The average AE is obtained by averaging the absolute error over the samples and over each DoF. 
The results indicate that the proposed TAP technique considerably outperforms other methods in terms of the minimizing the AE; hence, making it a crucial component of multi-service edge-intelligence.

\section{Open Research and Design Challenges}
\subsection{Edge-Intelligence-in-a-Box}
Real-time control applications have also started to emerge in the consumer sector. For instance, during the COVID-19 pandemic, teleoperation technology was used to remotely restock supermarket shelves\footnote{https://www.youtube.com/watch?v=UxWH5XAcFnM}. However, it relied on a specialized robot with complex hardware/software design. To fully unleash the benefits of teleoperation for businesses and industries and to bring it at scale, we introduce the concept of edge-intelligence-in-a-box. Such a box with predictive control capabilities will enable stable teleoperation for any application, utilizing off-the-shelf robotic hardware and haptic modules, and over any kind of connectivity interface, i.e., using both private and public 4G/5G or Wi-Fi. The box will deliver predicted command/feedback to its proximity operator/robot in a deterministic way using a TSN interface. Realizing such a box and conducting its trials is an open challenge which will be the focus of our future work. 

\subsection{Edge-Intelligence under Mobility}
User mobility associated with either master or slave domains directly affects the multi-service edge-intelligence framework. Realizing multi-service edge-intelligence under mobility becomes particularly challenging as it is not just a simple traffic path update problem, as in the case of conventional mobility solutions for MEC architectures. It requires extending the aforementioned co-design aspects (Section \ref{fw}) involving the source and target MEC hosts. Coordinated resource allocation, multi-connectivity, and federated ML techniques \cite{FL_intro} are the key enablers for designing a robust edge-intelligence framework under mobility, especially when the source and target MEC hosts have heterogeneous capabilities. 

\subsection{Co-design with Time-critical Computing}
Recent developments in the area of time-critical computing have led to a new class of hardware processors which are optimized for meeting the stringent temporal requirements of real-time applications. A prominent example is Intel\textsuperscript{\textcircled{R}} Time Coordinated Computing solution which fulfils hard real-time requirements in terms of jitter and latency. The time-critical computing paradigm enables industrial IoT devices to execute operations at fixed time scales. It also enables predictive engines to tightly couple model training and inferences phases. Extending multi-service edge-intelligence with co-design of TSN and time-critical computing is critical to unlocking the potential of time-critical control at scale, especially in safety-of-life applications.

\subsection{Multi-modal Prediction Challenge}
Perception in human-centric control applications is largely related integration of multiple sensory modalities (e.g., audio, visual, haptic, and tactile). Feedback in control applications typically involve multiplexing of different types of sensory information. However, different modalities have different tolerance levels to communication imperfections. This necessitates a multi-modal predictive framework that jointly predicts different types of sensory information while considering perceptual performance as well as optimized multiplexing strategies for the communication network.

\section{Concluding Remarks}
Multi-service edge-intelligence is a promising new paradigm to guarantee stability of time-critical control applications under a wide range of  wireless imperfections. This paper introduced its fundamental concept, which is based on tight coupling of MEC, ML techniques, and 5G RAN, along with some of the key system design aspects from a holistic perspective. Integrated MEC-5G systems for multi-service edge-intelligence heavily rely on the right deployment model with architectural, protocol, and radio resource allocation enhancements. Multi-service edge-intelligence can be realized as an edge-centric application via the ETSI-defined MEC reference architecture or in accordance with the O-RAN reference architecture. The paper also introduces a TAP technique which utilizes both short-term and long-term predictors to cope with the peculiarities of the wireless environment, and more importantly, to provide an application-agnostic approach to edge-intelligence functionality. Performance evaluation in a robotic manipulation scenario shows that the TAP technique outperforms conventional techniques in terms of overcoming wireless imperfections. To fully unleash the potential of multi-service edge-intelligence for time-critical control in industrial as well as consumer sectors, a number of challenges remain including edge-intelligence-in-a-box, operation under mobility, multi-modal predictive framework, and co-design involving real-time computing engines.

%\section*{Acknowledgment}

%This work has been partially supported by the ICT-ACROPOLIS Network of Excellence, FP7 project %no. 257626, www.ict-acropolis.eu.


%\section{Acknowledgement}
%The work presented in this paper is partly funded by the European Union’s Horizon 2020 research and innovation programme under grant agreement No 761745 and the Government of Taiwan. 
\bibliographystyle{IEEEtran}

\bibliography{IEEEabrv,EI_bib}
%\begin{thebibliography}{1}
%\end{thebibliography}

\begin{IEEEbiography}{Adnan Aijaz}
(M'14--SM'18)  studied telecommunications engineering at the King’s College London, U.K., where he received a Ph.D. degree in 2014 for research in wireless networks. He is currently the Programme Leader for Beyond 5G at the Bristol Research and Innovation Laboratory, Toshiba Research Europe Ltd., U.K. His recent research interests include 5G/6G wireless systems, Open RAN, time-sensitive networking, high-altitude platforms, and robotics and autonomous systems. 
\end{IEEEbiography}

\begin{IEEEbiography}{Nan Jiang} (M'20) was a Research Engineer with the Bristol Research and Innovation Laboratory, Toshiba Research Europe Ltd., U.K. from 2020 to 2021. He is currently a Research Engineer with the PHY Research \& Standards Lab, Samsung Research, Beijing. He was an Associate Professor with Beijing University of Posts and Telecommunications. He received the Ph.D. degree in Electronic Engineering from the Queen Mary University of London, UK., in 2020. He was a visiting researcher with the King's College London, U.K., in 2016 and 2018. He has served as a TPC Member for IEEE VTC'19, VTC'20, and VTC'22. His research interests include B5G, 6G, IoT, machine learning, and radio resource management.
\end{IEEEbiography}

\begin{IEEEbiography}{Aftab Khan} is the Distributed AI Programme Leader at the Bristol Research and Innovation Laboratory, Toshiba Europe Ltd., U.K. He received his Ph.D. in Machine Learning from the University of Surrey, U.K. (2013). His research agenda is mainly focused on distributed machine learning, AI-driven cyber security, computational behaviour analysis and pattern recognition. He has been involved in several EU and EPSRC projects (REPLICATE, SiDE, TEDDI, ACASVA) as well as industry led Innovate UK projects (SYNERGIA, CAVShield).
\end{IEEEbiography}

\end{document}